**Iron Nuclei with Energy of $10^{19}$ eV in Extragalactic Cosmic Rays Near Earth**

**A. V. Uryson***

*Lebedev Physical Institute, Russian Academy of Sciences, Moscow, 119991 Russia*

***e-mail: uryson@sci.lebedev.ru***

**Abstract**—The paper discusses the fraction of extragalactic cosmic iron nuclei with energies of $10^{19}$ eV that reached Earth from distant sources without fragmentation in intergalactic space. It is shown that the fraction of particles with energies $E = 10^{19}$ eV is significant (about 10%), even if the redshifts of the sources reach 0.2, and not only for nearby sources.

**Keywords:** cosmic rays, ultra-high energies, supermassive black holes, elemental composition, intergalactic space

## 1. INTRODUCTION

Cosmic rays (CRs) with energies $E > 10^{18}$ eV are likely extragalactic, as they are not confined by the magnetic fields of their host galaxies and can escape into intergalactic space.

The sources of extragalactic CRs are currently unclear. We hypothesize that CRs are accelerated near supermassive black holes (SMBHs). They have been detected at the centers of many galaxies, including our own Milky Way. According to current concepts, SMBHs exist in the core of virtually every galaxy [1]. CRs can be accelerated in the jet [2], accretion disk [3], and near the polar caps of SMBHs, where particles fall from the accretion disk [4, 5]; CRs can also gain energy in the ergosphere of a rotating SMBH [6, 7].

If this hypothesis is correct, virtually every galactic nucleus with an SMBH could be a source of extragalactic CRs.

The jet and accretion disk contain stellar matter, so the CR sources likely contain various elements, including iron nuclei.

Propagating in intergalactic space, CRs interact with background radiation—cosmic microwave, radio, and extragalactic light—in the following reactions:

$$A+\gamma \rightarrow A+e^{+}+e^{-} \quad (1)$$

(direct pair production, in CMS the threshold energy of 1 MeV),

$$A+\gamma \rightarrow A'+mN+n\pi \quad (2)$$

(photopion production, in CMS the threshold energy of 145 MeV),

$$A+\gamma \rightarrow A'+mN \quad (3)$$

(nuclear photodissosiation, in CMS the threshold energy of dozens of MeV).

As a result of these interactions, some of the primary nuclei and their fragments reach the array — the elemental composition of the CRs changes in intergalactic space. This is a consequence of the GZK effect [8, 9] as applied to nuclei. Photodisintegration of CR nuclei in intergalactic space was discussed previously in [10].

The elemental composition of extragalactic CRs is intensively studied due to the unresolved problem of their origin. CR particles are detected using ground based facilities, and the energy and mass number of a particle are determined indirectly — from the characteristics of the extensive air shower they cause. According to results obtained at the Telescope Array [11] and Pierre Auger [12] facilities, as well as at the Yakutsk facility [13], CRs with energies $E > 10^{19}$ eV contain iron nuclei.

It is clear that, if the UHECR sources are sufficiently close, the iron nuclei can reach Earth virtually without fragmentation (since particle free pass is too short to interact with background radiation). Then, based on the energy of the iron nuclei, it is possible to obtain distance constraints to their possible sources — whether they are located in the Local Group, the Local Supercluster, or beyond. Such analysis was conducted in [14]. Apparently, some fraction of iron nuclei with energies $E \geq 10^{19}$ eV may reach the Earth from distant sources. In this article, we discuss this fraction, although the flux of iron nuclei at energies of $10^{19}$ eV is insignificant — according to the results of [15], it is two orders of magnitude lower than the total CR flux.

## 2. MODELING AND ANALYSIS OF CALCULATION RESULTS

We calculated the fraction $R$ of particles with mass number $A$ = 56 (iron) reaching the facility from the possible source relative to all generated by iron nuclei secondary arriving particles with mass numbers from $A$ = 1 to $A$ = 56:

$$R = \text{(particles with } A=56\text{)/ (all arriving particles, } A=1\text{-}56\text{)}. \quad (4)$$

The value of $R$ was determined for different particle path lengths.

To find $R$, we simulated the propagation of iron nuclei in intergalactic space from the sources to Earth. The simulation was done under the following assumptions.

We assume that CRs are accelerated in point-like extragalactic sources with distances in the range $L \approx 2$–3500 Mpc.

Next, we assume that extragalactic CRs are iron nuclei ($A$ = 56). Of course, CRs contain both protons and nuclei of other elements, but we analyze possible sources of iron particles and therefore discuss the propagation in intergalactic space of iron nuclei only.

For distant sources, their evolution must be taken into account. The evolution of SMBHs is unclear. In our calculations, we adopted a form of source evolution that fully describes the data on extragalactic CRs when modeling CR propagation through intergalactic space. This is the evolution of one of the types of active galactic nucleus — BL Lac from [16].

Furthermore, according to the results of [16], data on extragalactic CRs can be described if, when modeling CR propagation, the injection spectrum in the sources is specified by a power-law $E^{-\gamma}$ with an exponent $\gamma$=2.2. Therefore, this injection spectrum was adopted in our model.

The maximum energy of iron nuclei in the source is of $10^{21}$ eV.

The minimum energy $E$min of a particle with charge $Z$, at which it can escape from its host galaxy, can be estimated from the relationship between $E_{min}$ and the radius of the particle's Larmor orbit $R_B$ in the galaxy magnetic field $B$, assuming that this radius exceeds at least the half-thickness of the galactic disk, $h_d$, $R_B \geq h_d$ (see, e.g., [17]):

$$E_{min} = 300 R_B Z B, \quad (5)$$

where the energy $E$min is measured in eV, the radius $R_B$ in cm, and the magnetic field $B$ in G.

Assuming that the sizes and magnetic fields of most host galaxies are approximately the same as those of our Galaxy, namely, the half-thickness of the disk $h_d \approx 100$ pc and the magnetic field $B \approx 10^{-6}$ G, we obtain the minimum energy of iron nuclei escaping from the host galaxy: $E_{min\ Fe} \approx 2.5 \times 10^{18}$ eV. Based on this, in the calculations, the minimum energy of iron nuclei in the source was assumed to be $5.6 \times 10^{18}$ eV.

In the model we use the characteristics of the background radio emission from [18, 19] and the extragalactic light from [20]. The cosmic microwave background (CMB) has Planck energy distribution with an average photon energy of $\varepsilon_r = 2.3 \times 10^{-4}$ eV and a photon density $n_r = 400$ cm$^{-3}$.

The photonuclear cross sections were taken from [21].

The simulations used the publically available code TransportCR code [22], which already included the source evolution form, background radiation parameters, and photonuclear cross sections from the papers above.

## 3. RESULTS AND DISCUSSION

The energy dependence $R(E)$ in the energy range $E \geq 6 \times 10^{18}$ eV for several source redshifts $z$ is shown in Fig. 1. It is clear that the fraction of iron nuclei reaching the array without fragmentation decreases with increasing nuclear energy and source redshift.

At $E = 10^{19}$ eV, the fraction of iron nuclei reaching the array is noticeable if the source redshift does not exceed 0.3: $R$ = 0.25, 0.09, and 0.02 at $z$ = 0.1, 0.2, and 0.3, respectively.

In cosmic ray physics, nuclei with charges Z ≥20 are grouped into the VH group - "very heavy nuclei." Therefore, we also determined the *R* value for the set of VH iron nucleus fragments that reached Earth. The energy spectra of the fragments do not coincide, and the intensities of the secondary components at a given energy differ by orders of magnitude.

We found that, for the chosen *z* values, the *R* value, starting from energies $E > 10^{19}$ eV, is $R \approx 1$, that means that iron nuclei with $E > 10^{19}$ eV from distant sources arrive at Earth as fragments with charges $Z \geq 20$. The integrated intensity of each secondary component is 56 times lower than the intensity of the iron nucleus. Fragment fluxes at energies of $10^{19}$ eV are very low due to the low flux of iron nuclei at this energy.

The model assumptions about the evolution of CR sources and their injection spectra do not affect the results. This is because the fraction of iron nuclei near Earth was calculated for sources with a given redshift *z*, and this fraction is determined solely by the propagation of particles in intergalactic space and is independent of the spatial density and luminosity of
the sources. For the same reason, the shape of the injection spectrum at the source adopted in the model does not affect the results.

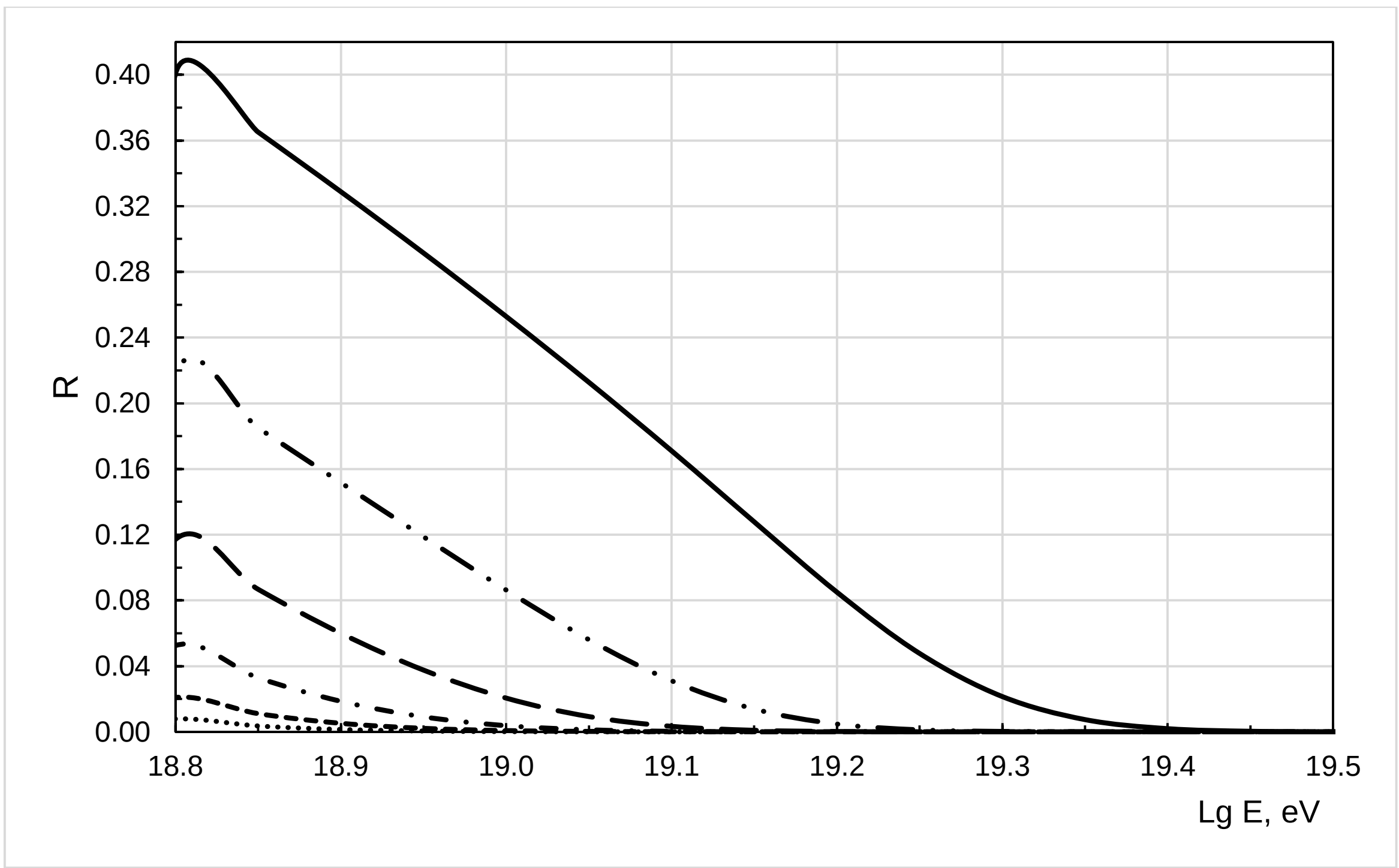

**Fig. 1.** *R* vs. the energy *E* of iron nuclei emitted by sources with fixed redshifts *z* = 0.1 (solid line), 0.2 (dash-and-two-dot line), 0.3 (dotted line), 0.4 (dash-and-one-dot line), 0.5 (short-dash-and-dot line), 0.6 and (dots).

## 4. CONCLUSIONS

In our model, extragalactic CRs are accelerated in the vicinity of SMBHs. It is currently generally accepted that a SMBH exists in the core of virtually every galaxy [1]. The elemental composition of CRs reflects the composition of the matter of stars captured by SMBHs, and therefore CRs contain iron nuclei.

The flux of iron nuclei at an energy of $10^{19}$ eV is low [15]. In our model, the fraction of iron nuclei with an energy $E = 10^{19}$ eV that reach the Earth constitutes a significant portion of this flux if the source redshift $z$ does not exceed 0.3: it is $R$ = 25, 9, and 2% for $z$ = 0.1, 0.2, and 0.3, respectively (which corresponds to source distances of 435, 850, and 1200 Mpc).

## ACKNOWLEDGMENTS

I am grateful to O.E. Kalashev for discussions of the TransportCR code and E.V. Bugaev, with whom the text of the article was discussed. I also thank the reviewer for discussions and comments.

FUNDING

This work was supported by ongoing institutional funding. No additional grants to carry out or direct this particular research were obtained.

CONFLICT OF INTEREST

The author of this work declares that he has no conflicts of interest.

*Translated by E. Chernokozhin*